# Antiferromagnetic dichroism and Davydov splitting of 3*d*-excitons in a complex multisublattice magnetoelectric CuB$_2$O$_4$

K. N. Boldyrev[1*], R. V. Pisarev[2], L. N. Bezmaternykh[3] and M. N. Popova[1]

[1]*Institute of Spectroscopy, Russian Academy of Sciences, 142190 Moscow, Troitsk, Russia*
[2] *Ioffe Physical-Technical Institute, Russian Academy of Sciences, 194021 St. Petersburg, Russia*
[3]*Kirensky Institute of Physics, Siberian Branch of the Russian Academy of Sciences, 660036 Krasnoyarsk, Russia*

The space and time symmetry breaking at magnetic phase transitions in multiferroics results in a number of strongly pronounced optical effects. Our high-resolution spectroscopic study of 3*d*- excitons in a complex multi-sublattice magnetoelectric CuB$_2$O$_4$ demonstrates that, among those, a large antiferromagnetic linear dichroism is observed which is highly sublattice-sensitive to subtle changes in the spin subsystems. We prove that the discovered linear dichroism is related microscopically to the magnetic Davydov splitting of the exciton states. We announce a novel magnetic phase transition and argue that an elliptical spiral structure rather than a simple circular helix is realized in the incommensurate phase, these findings being overlooked in previous studies by optical and other techniques. We claim that this spectroscopic method can be effectively applied to other materials for revealing hidden features of magnetic structures and phase transitions.

*Corresponding author.* E-mail address*: kn.boldyrev@gmail.com*

Phase transitions are widely spread phenomena in physics, chemistry, biology and even in social life. In the physics of solids, and in particular in periodic crystals, phase transitions are described in an elegant way using the symmetry arguments and the notion of the order parameter [1]. A strong burst in the development of physics of phase transitions took place when along with the space symmetry operations the symmetry of time reversal, or magnetic symmetry, was introduced [2,3]. This extension resulted in predictions and experimental observations of several new physical phenomena and new classes of materials, such as magneto-electrics, piezomagnetics, and multiferroics, with promising potentials for developing new technologies [4-9]. In particular, the space-time symmetry arguments opened new degrees of freedom in the studies of linear [10,11] and nonlinear [12] optical phenomena. However, beyond doubt, the abilities of optical spectroscopy in the studies of multiferroics remain far from being fully explored.

In this Letter, we demonstrate new capabilities offered by the high-resolution optical spectroscopy in the studies of multiferroics. This is done on the example of the copper metaborate CuB$_2$O$_4$, a compound exhibiting a unique combination of magnetic [13-15], magnetoelectric [16,17], linear and nonlinear optical [18-20] properties. This material is characterized by a complex phase diagram and a sequence of phase transitions with several features not observed before in any other transition-metal oxide. Magnetic control of the crystal chirality in CuB$_2$O$_4$ was recently reported on the basis of experiments with a circularly polarized light resonant with the 3*d* excitonic transitions [21] but later refuted on the basis of a theoretical symmetry analysis [22]. The latter publication was followed by a fierce "comment-reply" discussion raising questions of a fundamental importance [23,24], at the end of which participants remained with their own opinions [25,26]. Without any doubt, a comprehensive understanding of the nature of various optical effects observed in CuB$_2$O$_4$ is important, both from the point of view of basic physical concepts and in connection with a possible use of these effects in future optoelectronic devices based on remarkable properties of this compound [17].

Here we show that a large sublattice-sensitive linear dichroism (LD) emerges at an antiferromagnetic ordering of CuB$_2$O$_4$. We elucidate the nature of this LD attributing it to the magnetic Davydov splitting. On the example of CuB$_2$O$_4$, we show that the discovered LD can be used as a new method for revealing hidden features of magnetic structures and phase transitions in complex multi-sublattice multiferroics and magnetics. Our results allowed us to solve the problem concerning a possibility to control the crystal chirality by a magnetic field.

CuB$_2$O$_4$ crystallizes in the tetragonal noncentrosymmetric space group *I*-42*d* ( $D_{2d}^{12}$, Z=12), where the magnetic Cu$^{2+}$ ions (3*d$^9$*, *S*=1/2, *L*=0) occupy two distinct 4*b* and 8*d* square-coordinated positions with the $S_4$ and $C_2$ symmetry, respectively [27]. The presence of two substantially different positions for the copper ions leads to a number of interesting and unusual properties. The "strong" magnetic Cu(4*b*) subsystem spontaneously orders at the Néel temperature $T_N$=21 K into an antiferromagnetic commensurate structure with the spins lying in the easy (*xy*)-plane. The 4*b* spins are slightly canted due to the Dzyaloshinskii-Moriya interaction, with a weak ferromagnetic moment along the [110]-type axes [13-15]. Elastic [13] and inelastic [28] neutron scattering

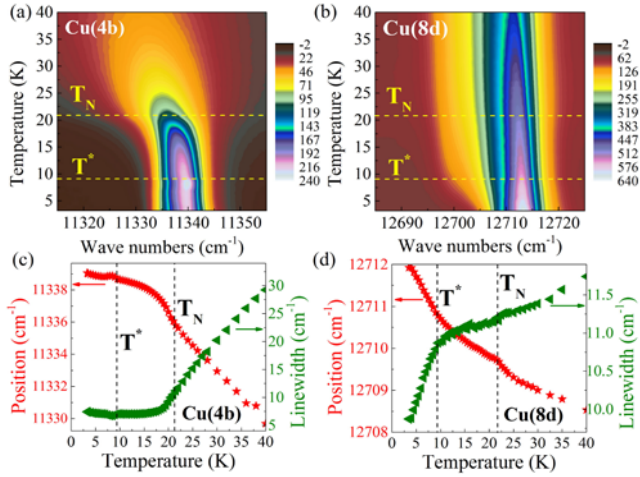

FIG. 1 (color online). Temperature behavior of the lowest-frequency Cu(4b) and Cu(8d) ZP lines. (a,b) Color-coded contour plots of the α-polarized absorption as a function of temperature vs wave number. The transition temperatures are marked as dotted lines. (c,d) Temperature dependences of the peak postions and line widths.

experiments revealed a quasi-one-dimensionality of the "weak" Cu(8d) magnetic subsystem. Whereas the Cu(4b) magnetic moments steadily grow below $T_N$ and reach the value $\mu_{Cu(b)} = 0.86\ \mu_B$ ($\mu_B$ is the Bohr magneton) at 12 K, the Cu(8d) ions at the same temperature acquire a small magnetic moment along the z-axis, $\mu_{Cu(d)} = 0.20\ \mu_B$; it starts to grow only below $T^* = 10$ K where the phase transition into an incommensurate helicoidal phase takes place, mounting to the value $\mu_{Cu(d)} = 0.54\ \mu_B$ at 2 K [4]. The Cu(8d) magnetic moments are not completely ordered and fluctuate even at the lowest temperatures.

Single crystals of $CuB_2O_4$ of good optical quality were grown as described previously [29]. Plane-parallel polished plates of (010) and (001) orientation with thickness between 50 and 100 μm were prepared from X-ray and optically oriented crystals. Optical transmission spectra, were measured for the range 11000-24000 cm$^{-1}$ (900 – 450 nm) with a resolution 0.8 cm$^{-1}$, using a Bruker IFS 125 HR Fourier spectrometer and a closed-cycle Cryomech ST403 cryostat at the temperatures between 2.5 and 300 K, stabilized with the precision up to 0.05 K. In addition to conventional measurements in the π- ($\mathbf{k}\perp z$, $\mathbf{E}(\omega)\|z$), σ- ($\mathbf{k}\perp z$, $\mathbf{E}(\omega)\perp z$), and α- ($\mathbf{k}\|z$, $\mathbf{E}(\omega)$, $\mathbf{H}(\omega)\perp z$) polarizations, we studied the angular dependence of the linear dichroic signal in the (xy)-plane, both without and in the magnetic field applied along the [110]- or [100]-type directions. The LD signal was found as the difference in the absorption coefficient for the light polarized along and perpendicular to some specified direction in the (xy)-plane. A magnetic field of up to 0.7 T was created by permanent neodymium-iron magnets carefully placed on both sides of a properly oriented sample.

Our measured absorption spectra in their general features are in agreement with earlier published data [10,11]. They demonstrate the six sharp zero-phonon (ZP) exciton lines originating from the d-d transitions of the $Cu^{2+}$ ions in both 4b and 8d positions, followed by broad intense bands with an unusually rich vibronic structure. However, our high-resolution measurements clearly revealed important details overlooked in the previous studies. Thus, the second 8d and 4b lines exhibit a doublet structure due to a low-symmetry component of the crystal field and the spin-orbit interaction, respectively (see the Supplemental Material A [30]). Figures 1 (a,b) and 1 (c,d) show the temperature behavior of the lowest-frequency Cu(4b) and Cu(8d) ZP lines, respectively. Both lines shift and narrow with lowering the temperature, but in a totally different way. Remarkably, all the 4b lines quickly broaden and disappear above $T_N$ whereas the 8d lines remain observable even above ~100 K. This unusual behavior deserves a more detailed investigation which is out of scope of this paper. The 4b lines demonstrate strikingly pronounced peculiarities at $T_N$ in both the position ν(T) and the line width δν(T) dependences. In contrast, the 8d lines remain almost unaltered at $T_N$, but shift and narrow dramatically at the temperature $T^*$ of the second phase transition.

These experimental data point to a distinct coupling of the 3d-excitons in both copper positions with the relevant magnetic phase transitions and can be naturally explained in the following way. At $T_N$, only the Cu(4b) magnetic moments undergo an ordering which results in shifting the energy levels and freezing the spin fluctuations present in the paramagnetic state of this copper subsystem. Correspondingly, the 4b ZP lines shift and narrow below $T_N$. At the phase transition at $T^*$, a similar scenario takes place for the Cu(8d) subsystem and the relevant 8d spectral lines. Both copper subsystems, Cu(4b) and Cu(8d), are almost independent and therefore the ZP lines of each subsystem demonstrate, predominantly, magnetic properties of the relevant particular subsystem. Nevertheless, weak but well pronounced features in the 8d lines are observed also at $T_N$, as well as the 4b lines show marked frequency changes at $T^*$ (see Fig. 1). This behavior of the 4b and 8d excitons clearly demonstrates an existence of a mutual coupling between them. Thus, this coupling being too weak to be observable in magnetic measurements [4] finds a clear confirmation in the behavior of the exciton lines.

Next, we discuss important polarization properties of the ZP lines for the case of light propagating along the tetragonal z-axis, $\mathbf{k}\|z$. Above $T_N$, the macroscopic magnetic symmetry of $CuB_2O_4$ is $\bar{4}2m\underline{1}$ and the (xy)-plane is magnetically and optically isotropic. Therefore, the absorption for $\mathbf{k}\|z$ is also isotropic and does not depend on the orientation of $\mathbf{E}(\omega)$ with respect to the x- or y-axes. However, right below $T_N$ the spins of the 4b subsystem order in the (xy)-plane along the [110] or [$\bar{1}$10] type axes reducing the magnetic point group to $m\underline{m}2$ in which the

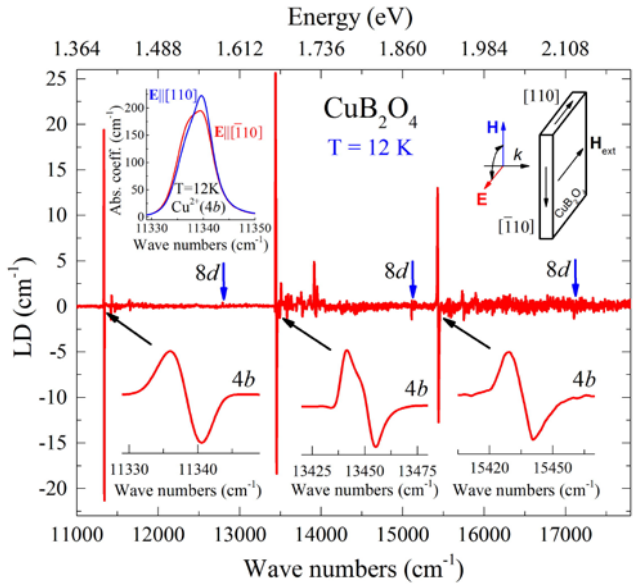

FIG. 2 (color online). (a) Spectrum of the linear dichroism (a difference between the spectra registered in $k\|z$, $E\|[110]$ and $k\|z$, $E\|[\bar{1}10]$ polarizations) of $CuB_2O_4$ in the commensurate magnetic phase at 12 K. Positions of ZP lines originating from the $Cu^{2+}(4b)$ and $Cu^{2+}(8d)$ positions are marked by arrows, in accordance with Ref. 19. Lower insets demonstrate the shapes of dichroic signals observed for the three $Cu^{2+}(4b)$ ZP lines. Upper left inset shows the first $Cu^{2+}(4b)$ ZP line registered in two mutually perpendicular polarizations, $E\|[110]$ and $E\|[\bar{1}10]$, both for $k\|z$, evidencing the Davydov doublet and explaining the observed LD (lower left inset). Upper right inset displays geometry of the experiment. Pay attention that the $Cu^{2+}(8d)$ ZP lines do not exhibit LD.

fourfold axis $\bar{4}$ is absent. This process results in a breaking of the macroscopic magnetic symmetry and, consequently, the optical isotropy in the $(xy)$-plane must be destroyed. The magnetic point group $mm2$ allows the magnetic linear dichroism which is the difference in the absorption of light polarized along and perpendicular to the spin directions [3,22,24,25]. We note that this point group also allows a weak ferromagnetism, which was observed experimentally in the $(xy)$-plane at 90° with respect to the $4b$ antiferromagnetic spins [6]. We note that the magnetoelectric and the piezomagnetic effects are also allowed in this point group [2]. In the case of the $B\|x$ and $B\|y$ geometries, the magnetic point groups are $2\underline{22}$ and $\underline{2}2\underline{2}$, respectively and, again, the LD is, allowed [3,24].

Indeed, an appreciable linear dichroism for all three $4b$ ZP lines is observed below $T_N$ as shown in Fig. 2. One has to keep in mind the existence of two kinds of antiferromagnetic domains characterized by the $4b$ copper spins along the [110] or [$\bar{1}$10] directions and giving opposite contributions into the observed effect. Evidently, the magnitude of LD can be considerably enhanced by applying a magnetic field in the $(xy)$-plane which favors one type of domains at the expense of the other making a single-domain sample. Really, we observed experimentally a strong enhancement of the LD by applying a magnetic field of 0.7 T in the $(xy)$-plane, $B\|[110]$. We also studied the $B\|x$ and $B\|y$ geometries and found a strong LD in this case. As for the $8d$ ZP lines, none of them shows any dichroism in this geometry, which is understandable in view of the alignment of $Cu(8d)$ magnetic moments predominantly along the $z$-axis.

It is worth noting that the shape of a dichroic signal (Fig. 2) is similar to that observed in Ref. [21], whereas its value exceeds that of Ref. [21] by about two orders of magnitude. We can suggest that the authors of Ref. [21] also observed the LD signal, due to a small linearly polarized component in their circularly polarized light.

After unambiguously establishing the presence of a strong antiferromagnetic linear dichroism, the important question arises what is its *microscopic* origin, in particular, for the first ZP line which corresponds to the optical transition $(x^2-y^2)\rightarrow(xy)$ between two non-degenerate orbital singlet states (the Supplemental Fig. S1a). To understand this, we pay attention to the presence of two crystallographically equivalent $Cu(4b)$ ions in the *primitive* cell of $CuB_2O_4$. General arguments show that in this case one might expect a Davydov excitonic doublet which depends on the magnetic structure [31,32]. (In the Supplemental Material B [30], this is explained in detail for the magnetic structure realized in the temperature interval $T^* < T < T_N$). And indeed, below $T_N$ two orthogonally polarized components, at 11336 and 11340 cm$^{-1}$, are resolved within the first $4b$ ZP line (Inset of Fig. 2).

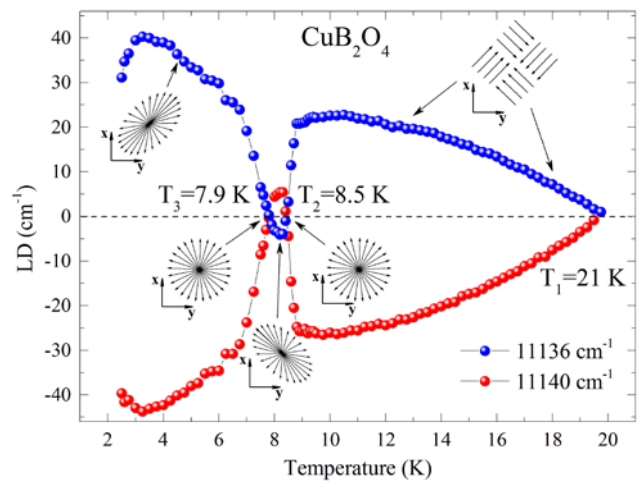

FIG. 3 (color online). Temperature dependence of LD (for the light polarized along and perpendicular to the [110] direction) at 11140 cm$^{-1}$ (red balls) and 11136 cm$^{-1}$ (blue balls). A splitting of the phase transition at $T^*$ into two transitions (at $T_2$ and $T_3$) is evident. Arrows schematically demonstrate proposed magnetic structures.

The antiferromagnetic-order-induced linear dichroism discovered in our study opens an opportunity to probe different magnetic phases of a rich B-T phase diagram of $CuB_2O_4$. Figure 3 shows dichroic signals at the wavelengths of the two Davydov components of the first $4b$ ZP line as functions of temperature, in a zero external

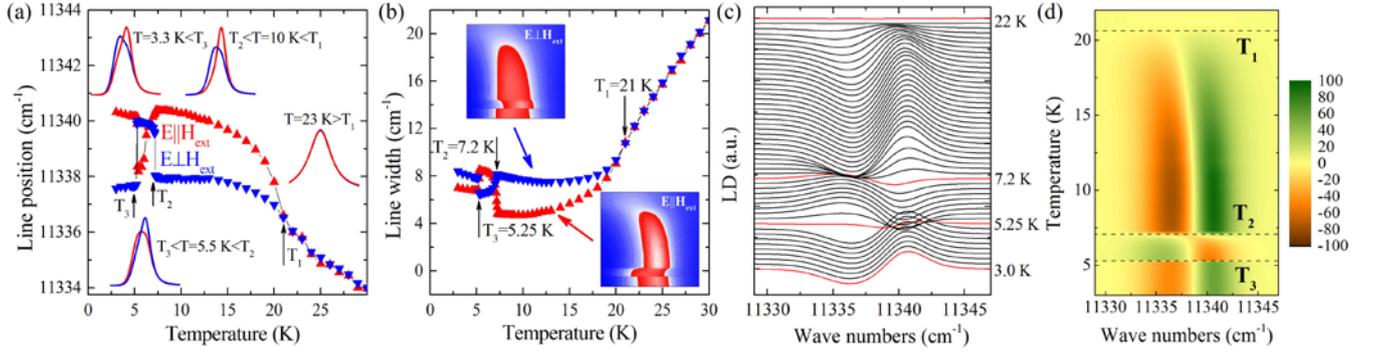

FIG. 4 (color online). Spectral signatures of phase transitions in CuB$_2$O$_4$ placed into an external magnetic field $B_{ext}$=0.7 T applied along the [110] direction. (a) Peak positions for the first 4$b$ ZP line in the $E\|B_{ext}$ (red symbols) and $E\perp B_{ext}$ (blue symbols) polarized spectra vs temperature. Insets show the spectral shapes of the $E\|B_{ext}$ (red curves) and $E\perp B_{ext}$ (blue curves) polarized first 4$b$ ZP line in different temperature regions between the phase transitions. (b) Widths (FWHM) of the first 4$b$ ZP line in the $E\|B_{ext}$ (red symbols) and $E\perp B_{ext}$ (blue symbols) polarized spectra vs temperature. Insets show color-coded contour plots of the $E\|B_{ext}$ and $E\perp B_{ext}$ polarized absorption as a function of temperature vs wave number. (c, d) LD spectra in the region of the first 4$b$ ZP line of CuB$_2$O$_4$ at temperatures below $T_N$, (for the light polarized along and perpendicular to the [110] direction; $\mathbf{k}\|z$), displayed as (c) shifted plots taken with steps from 0.5 to 0.05 K and (d) color-coded contour plots.

magnetic field. A splitting of the phase transition at $T^*$ into two transitions ($T_2$ = 8.5 K and $T_3$ = 7.9 K) is clearly seen, with an intriguing change of the LD sign between $T_2$ and $T_3$. At further lowering the temperature below $T_3$, the LD grows steadily. One more upturn observed at ~3 K is, evidently, connected with low-temperature phase transitions below 2 K [33]. Two phase transitions in the vicinity of $T^*$ are clearly observed in our LD spectra in the case of an applied magnetic field $B\|[110]$ which moves both $T_2$ and $T_3$ temperatures to lower values while the ($T_2$ - $T_3$) difference grows (Fig. 4). This behavior is in agreement with Ref. [34] which reported an existence of two different incommensurate magnetic phases in CuB$_2$O$_4$ at $T<T^*$, under applied magnetic field B>1.2 T. The nature of these phases was not specified in Ref. [34]. Our data show that, under an external magnetic field, the phase transition at $T_2$ is of a second order but that at $T_3$ is a first-order one (Fig. 4b,d).

All the available literature data point to the incommensurability of magnetic structures in the left-side part of the phase diagram of CuB$_2$O$_4$. A simple spin helix magnetic structure was suggested in a number of works, with the propagation vector along the tetragonal $z$-axis [14,15,33]. However, this picture is not compatible with our experimental magnetic LD data because in the case of a simple spin helix the LD would obviously vanish by symmetry arguments. Experimentally, LD vanishes only at two temperature values, $T_2$ and $T_3$, see Figs. 3 and 4. We may confidently assume that not a simple helix but some kind of elliptical helicoidal spin structures is realized. An elliptical structure was suggested in the theoretical work [35] but no experimental confirmation was found before. A sign change of the LD at $T_2$ and, once more, at $T_3$, can be interpreted as a reorientation of the long axis of the spin ellipse at $T_2$ and then one more reversal at $T_3$. Detailed measurements of the angular dependences of the dichroic signal at different temperatures show a smooth rotation of the long axis of a spin ellipse below $T_2$. Such a plethora of magnetic structures and phase transitions in CuB$_2$O$_4$ evidently takes place due to multiple frustrated and nonfrustrated antisymmetric exchange interactions within and between the magnetic 4$b$ and 8$d$ subsystems [35].

Thus, the present high-resolution spectroscopic study of exciton lines at both 4$b$ and 8$d$ magnetic sites of CuB$_2$O$_4$ evidences a pronounced coupling between optical electronic transitions and changes in magnetic subsystems in this multisublattice compound. A weak interaction between 4$b$ and 8$d$ magnetic subsystems was observed which was not noticed in numerous previous studies by optical and other methods. Our study has allowed us to resolve the magnetic Davydov splitting of the 3$d$ copper excitons in a complex magnetoelectric antiferromagnet CuB$_2$O$_4$ and, as a consequence, to observe a well pronounced sublattice-sensitive antiferromagnetic linear dichroism in the crystallographically isotropic ($xy$)-plane of this tetragonal crystal. We have shown that the discovered linear dichroism can serve as a highly sensitive tool for probing magnetic phase transitions and magnetic structures. In particular, we have found a new magnetic phase transition in CuB$_2$O$_4$, overlooked in previous optical and non-optical studies. Moreover, we strongly argue against a simple spin helix structure in the incommensurate magnetic phases of CuB$_2$O$_4$ and suggest an elliptical spin structure which rotates when the temperature is changed. No doubt that such an approach based on general symmetry principles can be applied to the studies of magnetic phase transitions and structures in other materials. In addition, our findings are against the claim of ability of a magnetic field to control the crystal chirality (Refs. [21,23]). We put forward an alternative explanation of the experimental results reported in Ref. [21] and confirm the correctness of the magnetic symmetry arguments about magneto-optical effects in magnetoelectric $CuB_2O_4$ (Refs. [22,24]).

The authors acknowledge financial support from the Russian Academy of Sciences under the Programs for Basic Research. K.N.B. and M.N.P. acknowledge the support from the Russian Science Foundation. R.V.P. acknowledges the support from the Russian Ministry of Education and Science under the Grant #14.B25.0031.25, and a partial support from the Russian Foundation for Basic Research, Project #12-02-00130a.**References:**

[1] L. D. Landau and E. M. Lifshitz, *Statistical Physics*, Pergamon (1980).

[2] L. D. Landau and E. M. Lifshitz, *Electrodynamics of continuous media.* Pergamon (1984).

[3] R. R. Birss, R. R. *Symmetry and Magnetism,* North Holland, Amsterdam (1967).

[4] M. Fiebig, *J. Phys. D* **38**, R123-R152 (2005).

[5] W. Eerenstein, N. D. Mathur, and J. F. Scott, *Nature* **442**, 759-765 (2006).

[6] D. I. Khomskii, *J. Magn. Magn. Mater.* **306**, 1-8 (2006).

[7] S.-W. Cheong and M. Mostovoy, *Nature Materials* **6**, 13-20 (2007).

[8] Y. Tokura and S. Seki, *Adv. Mater.* **22**, 1554-1565 (2010).

[9] A. P. Pyatakov and A. K. Zvezdin, *Physics-Uspekhi* **55**, 557-581 (2012).

[10] R. V. Pisarev, *Ferroelectrics* **162**, 191-209 (1994).

[11] T. Arima, *J. Phys. Condens. Matter* **20**, 434211 (2008).

[12] M. Fiebig, V. V. Pavlov, and R. V. Pisarev, *J. Opt. Soc. Amer. B* **22**, 96-118 (2005).

[13] G. Petrakovskii, D. Velikanov, A. Vorotinov, A. Balaev, K. Sablina, A. Amato, B. Roessli, J. Schefer, and U. Staub, J. Magn. Magn. Mater. **205**, 105 (1999).

[14] B. Roessli, J. Schefer, G. Petrakovskii, B. Ouladdiaf, M. Boehm, U. Staub, A. Vorotinov, and L. Bezmaternykh, Phys. Rev. Lett. **86**, 1885 (2001).

[15] M. Boehm, B. Roessli, J. Schefer, A. Wills, B. Ouladdiaf, E. Lelievre-Berna, U. Staub, and G. Petrakovskii, Phys. Rev. B **68**, 024405 (2003).

[16] N. D. Khanh, N. Abe, K. Kubo, M. Akaki, M. Tokunaga, T. Sasaki, and T. Arima, Phys. Rev. B **87**, 184416 (2013).

[17] M. Saito, K. Ishikawa, S. Konno, K. Taniguchi, and T. Arima, Nature Materials **8**, 634 (2009).

[18] R. V. Pisarev, I. Sänger, G. A. Petrakovskii, and M. Fiebig, Phys. Rev. Lett. **93**, 037204 (2004).

[19] R. V. Pisarev, A. M. Kalashnikova, O. Schöps, and L. N. Bezmaternykh, Phys. Rev. B **84**, 075160 (2011).

[20] R. V. Pisarev, K. N. Boldyrev, M. N. Popova, A. N. Smirnov, V. Yu. Davydov, L. N. Bezmaternykh, M. B. Smirnov, V. Yu. Kazimirov, Phys. Rev. B **88**, 024301 (2013).

[21] M. Saito, K. Ishikawa, K. Taniguchi, and T. Arima, Phys. Rev. Lett. **101**, 117402 (2008); M. Saito, K. Taniguchi, and T. Arima, J. Phys. Soc. Jap. **77**, 013705 (2008).

[22] S. W. Lovesey and U. Staub, J. Phys.: Condens. Matter **21**, 142201 (2009).

[23] T. Arima and M. Saito, J. Phys.: Condens.Matter **21**, 498001 (2009).

[24] S. W. Lovesey and U. Staub, J. Phys.: Condens. Matter **21**, 498002 (2009).

[25] S. W. Lovesey and E. Balcar, Physica Scripta **81**, 065703 (2010).

[26] N. D. Khanh, N. Abe, K. Kubo, M. Akaki, M. Tokunaga, T. Sasaki, and T. Arima, Phys. Rev. B **87**, 184416 (2013).

[27] M. Martinez-Ripoll, S. Martinez-Carrera, S. Garcia-Blanco, Acta Crystallogr. B **27**, 677 (1971).

[28] S. Martynov, G. Petrakovskii, and B. Roessli, J. Magn. Magn. Mater. **269**, 106 (2004).

[29] K. S. Aleksandrov, B. P. Sorokin, D. A. Glushkov, L. N. Bezmaternykh, S. I. Burkov, and S. V Belushchenko,. Phys. Sol. State **45**, 42 (2003).

[30] See Supplemental Material at [URL will be inserted by publisher].

[31] A. S. Davydov, Theory of molecular excitons. Plenum Press, New York (1971).

[32] R. Loudon, Advances in Physics **17**, 243 (1968).

[33] A. I. Pankrats, G. A. Petrakovskii, M. A. Popov, et al., JETP Lett. **78**, 569-573 (2003).

[34] Y. Kousaka, S. Yano, M. Nishi, K. Hirota, J. Akimitsu, J. Phys. Chem. Solids, **68**, 2170 (2007).

[35] S. N. Martynov, JETP Lett. **90**, 55 (2009).5